\documentclass[pra,twocolumn,showpacs,groupedaddress,nofootinbib]{revtex4-1}
\usepackage{amsmath,amssymb,amsfonts}
\usepackage{graphicx}
\usepackage{txfonts}
\usepackage{setspace}
\usepackage{color}
\usepackage{mathtools}
\usepackage{upgreek}
\usepackage{url}

\usepackage[unicode,breaklinks]{hyperref}

\hypersetup{
    unicode=true,
    a4paper=true,
    plainpages=false,
    pdfauthor={},
    colorlinks=true,
    linkcolor=blue,
    citecolor=blue,
    filecolor=black,
    urlcolor=blue
}
\newcommand{\ket}[1]{|#1\rangle}

\begin{document}

\title{Measuring the disorder of vortex lattices in a Bose-Einstein condensate}

\author{A.~Rakonjac}
\author{A.~L.~Marchant}
\author{T.~P.~Billam}
\author{J.~L.~Helm}
\author{M.~M.~H.~Yu}
\author{S.~A.~Gardiner}
\author{S.~L.~Cornish}

\affiliation{Joint Quantum Centre (JQC) Durham - Newcastle, Department of Physics, Durham University, Durham DH1 3LE, United Kingdom}

\date{\today}

\begin{abstract}

We report observations of the formation and subsequent decay of a vortex lattice in a Bose-Einstein condensate confined in a hybrid optical-magnetic trap. Vortices are induced by rotating the anharmonic magnetic potential that provides confinement in the horizontal plane. We present simple numerical techniques based on image analysis to detect vortices and analyze their distributions. We use these methods to quantify the amount of order present in the vortex distribution as it transitions from a disordered array to the energetically favorable ordered lattice.

\end{abstract}


\maketitle


\section{Introduction}

Rotating Bose-Einstein condensates (BECs) provide a highly controllable and versatile experimental platform for the study of fundamental aspects of superfluidity and turbulence \cite{Fetter1966, Anderson2010}. Following the creation of quantized vortices \cite{Matthews1999, Madison2000}, experiments have studied the formation of large-scale vortex lattices \cite{AboShaeer2001, Engels2002}, the role of vortices in the Berezinskii-Kosterlitz-Thouless phase transition \cite{Hadzibabic2006}, and their appearance as topological defects in the Kibble-Zurek mechanism \cite{Weiler2008}. More recently, advances in real-time vortex imaging \cite{Freilich2010, Wilson2015} and experimental control of vortex creation \cite{Wilson2013} have accompanied a growing interest in quantized vortex turbulence, from both experimental \cite{Neely2013, Kwon2014, Kwon2015} and theoretical \cite{Nowak2012,
Schole2012, Reeves2013, Billam2014, Simula2014, Stagg2015} perspectives.

Nucleating vortices in a BEC is possible using a variety of techniques that can impart angular momentum to the system, such as internal state manipulation \cite{Matthews1999}, stirring using a laser beam \cite{Madison2000}, dynamically manipulating the trapping potential \cite{Hodby2001, Kang2015}, and topological phase imprinting \cite{Leanhardt2002}. Following a period of induced rotation, a single-component BEC is in a highly non-equilibrium state as vortices nucleate. Like-signed vortices subsequently crystallize into a regular lattice, usually of a triangular geometry \cite{Parker2005}. Other lattice geometries are also possible. For example, square lattices can form in the presence of dipolar interactions \cite{Zhang2005} or in two-components gases \cite{Schweikhard2004}. This emerging order is indicative of the system relaxing to a lowest-energy equilibrium state. The time scale for the lattice to form can be either dependent \cite{Madison2001} or independent \cite{AboShaeer2002} of temperature, depending on the particular stirring mechanism used \cite{Lobo2004}, with dissipation mechanisms involving the thermal component playing a role in the former case and dynamic instability in the latter. While much theoretical attention has been given to vortex lattice formation in stirred two-dimensional (2D) systems \cite{Campbell1979, Aftalion2001, Tsubota2002, Kasamatsu2003, Lundh2003, Parker2005, Wright2008}, a comparatively smaller body of work exists for the three-dimensional (3D) regime \cite{Lobo2004, Kasamatsu2005}. A careful study, both experimentally and theoretically, of the regime in-between vortex nucleation and crystallization of the lattice may elucidate the mechanisms of energy dissipation in both 2D and 3D geometries.

There is a stark qualitative difference between the disordered collection of vortices present initially in a rapidly rotating BEC and the ordered lattice that subsequently forms as the system relaxes into a lower-energy configuration. The number of vortices present along with their configuration characterize the rotational equilibrium states of an irrotational fluid which minimize the free energy of the system. Theoretically, one can quantify the order of a vortex lattice by comparing the free energy of a given configuration of vortices to the free energy of the lowest-energy equilibrium state \cite{Campbell1979, Aftalion2001, Aftalion2012}. Previously, attempts have been made to quantify order in vortex lattices by examining pair correlations \cite{Engels2002, Bradley2008}, although comparisons between ordered and disordered vortex distributions were only made qualitatively. In superconductor vortex lattices, which also exhibit a triangular geometry \cite{Abrikosov1957}, disorder has been quantified by calculating translational and orientational correlations, as well as by examining clusters of lattice defects \cite{Zehetmayer2015}. 

In this work, we study the growth and decay of vortex lattices in a $^{87}$Rb BEC in a 3D trapping geometry. To rotate the BEC, we use the method of Kang \textit{et al.}~\cite{Kang2015} and gently revolve an anharmonic magnetic quadrupole potential which contributes to the confinement of the condensate. The anharmonicity of the potential couples the center-of-mass motion of the condensate to its internal motion and thereby imparts angular momentum, akin to how a wine connoisseur swirls wine in a glass prior to tasting. We describe an automated vortex detection algorithm suitable for use with large data sets. We focus particularly on the systematic analysis of lattice disorder using simple numerical techniques that provide a heuristic measure of lattice energy via two different metrics. The first metric is based on the single distance scale present in an equilateral triangular geometry, and the second is derived from fitting a triangular lattice to a vortex distribution. Using these metrics, we track the evolution of the system from a disordered arrangement of vortices to an ordered lattice, observing a clear transition to an ordered state.


\section{Experimental Overview}

\begin{figure}[t!]
	\centering
	\includegraphics{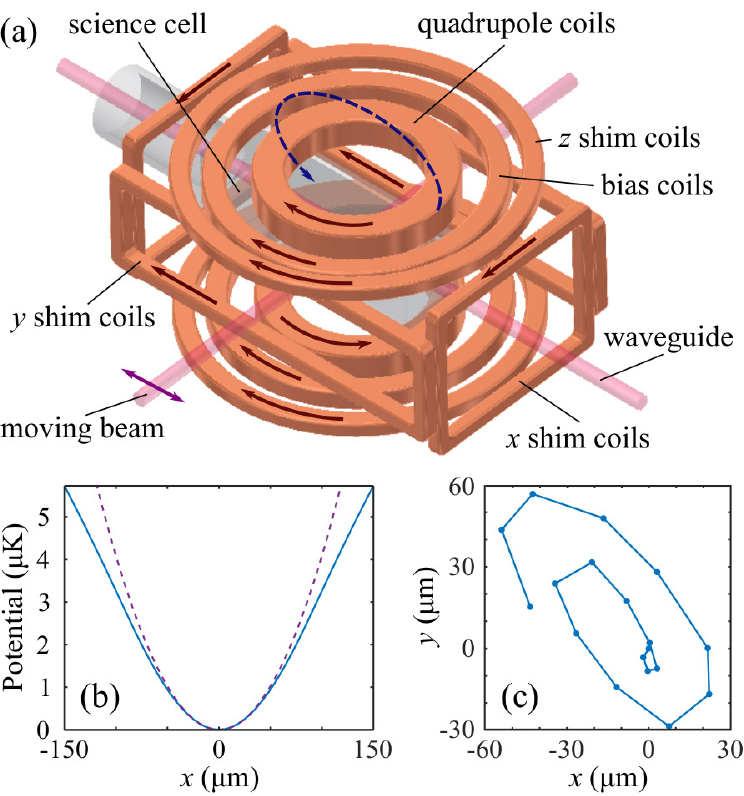}
	\caption{(Color online) (a) The experimental setup, showing the science cell, the four sets of coils relevant to inducing rotation, and the optical dipole trapping beams. The moving beam and waveguide cross at 90$^\circ$. The solid arrows on the coils indicate direction of current flow, while the dashed blue line illustrates the rotating magnetic field. The combined optical and magnetic potential (solid blue line) is plotted in (b). For comparison, a harmonic fit to the potential is also shown (purple dashed line). The position of the BEC as it spirals out from its initial position over the 50~ms of driving time is shown in (c). The positions are measured at 2.5~ms intervals after 12~ms of time of flight.}
	\label{fig:experimentschematic}
\end{figure}

Details of our experimental apparatus have been described elsewhere \cite{Handel2012}. $^{87}$Rb atoms are  initially cooled in a magneto-optical trap before being optically pumped into the $\ket{F = 1, m_F = -1}$ state and transferred to the science cell (shown in Fig.~\ref{fig:experimentschematic}) using a mechanical magnetic transport scheme \cite{Lewandowski2003}. Here the atoms are loaded into a magnetic quadrupole trap where they undergo forced radio-frequency evaporation until the lifetime becomes limited by Majorana spin flips. At this point, we transfer the atoms into a hybrid optical and magnetic trap \cite{Lin2009} by linearly ramping the quadrupole gradient from 180 down to 30~G/cm.  

The hybrid trap consists of a crossed optical dipole trap positioned $\sim 150~\mu$m below the field zero of the weak quadrupole potential. The dipole trap is formed from a versatile \textit{moving beam} which crosses at 90$^{\circ}$ with a second beam, referred to as the \textit{waveguide}, as shown in Fig.~\ref{fig:experimentschematic}. The trapping light is provided by a multi-mode 1070~nm fiber laser (YLR-50-1070-LP from IPG Photonics), with the moving and waveguide beams having waist sizes of 68(3) and 112(3)~${\mu}$m, respectively. The position of the moving beam is controlled using an acousto-optic deflector, which is capable of deflecting the beam up to 3~mm with sub-micron precision at  modulation frequencies greater than 1~MHz. This allows us to generate time-averaged traps by rapidly dithering the beam position \cite{Schnelle2008, Henderson2009}, which is utilized in the experiment described here, as well as to transport the atoms over several millimeters with high precision \cite{Roberts2014}. 

In addition to the magnetic quadrupole coils, our apparatus contains three sets of shim coils used to apply small ($<10$~G) magnetic fields allowing the position of the magnetic field zero to be precisely controlled. The positioning of the crossed dipole trap relative to the field zero is crucial for the rotation experiments described here as this determines the frequency and anharmonicity of the trapping potential. The set of bias coils also shown in Fig.~\ref{fig:experimentschematic} can provide larger fields and are used in levitated time of flight. 

The hybrid trap is initially loaded using beam powers of 7.8 and 5.6~W for the moving and waveguide beams, respectively. Note that this does not result in an equal contribution to the trap depth from the two beams; the lower power used in the waveguide is due to a technical constraint in our setup. We begin optical evaporation by first linearly reducing the power in the moving beam until the trap depth it provides matches that of the waveguide in the horizontal plane. We then simultaneously linearly reduce the power of both beams to drive evaporation in stages interspersed with periods of constant power, where the atoms undergo passive evaporation. This produces BECs of 8.3(1)~$\times~10^5$ atoms in a final trapping geometry of $\omega_{x,y,z} \approx 2 \pi \times (44,30,42)$~Hz.

Several groups have reported intensity-dependent optical pumping from $\ket{F = 1}$ to $\ket{F = 2}$ in $^{87}$Rb by trapping light generated by multi-mode fiber lasers \cite{Lauber2011, Kumar2012, Hung2015}. We have also observed this. Lauber \textit{et al.}~\cite{Lauber2011}, who use the same model of laser, found that the effect was minimized below an intensity on the order of 1~kW/mm$^2$. This corresponds to the maximum intensity used in our experiment, and we consequently see only minimal transfer to $\ket{F = 2}$ of approximately 1$\%$. Despite the associated heating and atom loss from our desired state, we are able to create BECs with lifetimes of many seconds, giving us ample time to perform the experiments described here.

To study vortex dynamics, we must first transfer the BEC into a trap where the confinement in the $xy$ plane is dominated by an anharmonic potential \cite{Marzlin1998}. To achieve this, we remove the waveguide beam and dither the position of the moving beam to reduce the horizontal confinement from the dipole trap. The optical potential now only provides confinement in the $z$ direction with confinement in the $xy$ plane being provided by the magnetic quadrupole potential, 
\begin{equation}
U(r) = \mu B' \sqrt{\frac{x^2}{4}+\frac{y^2}{4}+z_{\mathrm{offset}}^2},
\label{eq:MagPot}
\end{equation}
where $\mu$ is the magnetic moment of the atoms, $B'$ is the magnetic field gradient along $z$, and $z_{\mathrm{offset}}$ is the position of field zero with respect to the dipole trap. At the initial field zero position $z_{\mathrm{offset}}=150~\mu$m, this potential is approximately harmonic over the size of the atomic cloud with a frequency given by $\omega_{x,y} = \sqrt{\mu B'/ 4 m z_{\rm{offset}}} \simeq 2 \pi \times 20$~Hz. By bringing the field zero much closer to the dipole beam (i.e., reducing $z_{\mathrm{offset}}$) the range over which the potential is harmonic is reduced and the trapping in the horizontal plane becomes tighter and strongly anharmonic as $z_{\mathrm{offset}}$ becomes comparable to the cloud size. The final potential in the $x$ direction is shown in Fig.~\ref{fig:experimentschematic}(b). 

\begin{figure*}
	\centering
	\includegraphics[width=\textwidth]{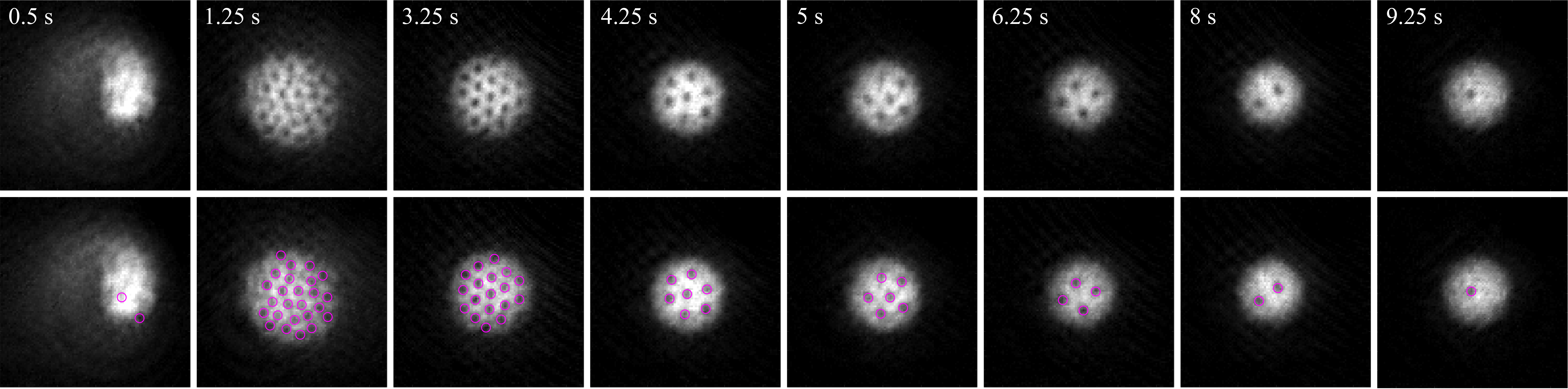}
	\caption{(Color online) Formation and decay of a vortex lattice at successive hold times following the rotation procedure, imaged after 42~ms of time of flight. Each image is 510~$\times$~510~$\mu$m$^2$. At 0.5~s, the cloud is still undergoing obvious center-of-mass motion. A disordered distribution of vortices appears, which crystallizes into a triangular lattice and subsequently decays. The bottom row depicts the same images as the top row, but with circles showing vortices detected by the vortex detection algorithm. The algorithm is most successful at detecting vortices in images with good contrast. Performance is worse for low contrast images such as the image at 1.25~s.}
	\label{fig:latticeevolution}
\end{figure*}

To create this new trap, we first ramp up the moving beam power over 50~ms to compensate for the reduction in trap depth in the time-averaged potential. We then reduce the waveguide power to zero over 150~ms. To broaden the moving beam potential, we apply a 5~kHz modulation, increasing the amplitude linearly over 500~ms to its final value of 70~${\mu}$m. The quadrupole field zero is then shifted vertically towards the atoms using the $z$ shim coils. At the end of a 50~ms linear ramp, the field zero is positioned 28(1)~$\mu$m above the dipole trapping beam. We infer this distance from trapping frequency measurements. Although this potential is anharmonic, for very small horizontal displacements, effective trapping frequencies can be measured from small amplitude trap oscillations. We measure these frequencies to be $\omega_{x,y,z}=2\pi~\times~[43(5), 46(4), 35(2)]$~Hz. The Thomas-Fermi radii of the condensate in this trap are approximately 27~$\mu$m horizontally and 33~$\mu$m vertically.
 
To impart rotation into the system, we displace the magnetic field zero, thus shifting the trap center. The position of the field zero is rotated by sinusoidally modulating the current in the $x$ and $y$ shim coils at 37~Hz, close to the trapping frequencies, for 50~ms using two phase-synchronized channels of a function generator $\pi/2$ out of phase. This is a delicate process; the amplitude of the resultant quadrupole field displacement is only $\sim$5$\mu$m, and it is crucial that any stray magnetic fields are completely nulled. The resulting center-of-mass motion of the BEC is mapped out in Fig.~\ref{fig:experimentschematic}(c). It is elliptical due to a small fixed deviation in phase from $\pi/2$ between the $x$ and $y$ modulation signals.

Once the trap rotation is complete, we hold the atoms in the now static trap for varying lengths of time to allow the system to evolve and equilibrate. We release the atoms from the trap and, following 42~ms of time of flight, image the cloud with a magnification of 3.1 using a resonant probe beam propagating along $z$. For 40~ms of the time of flight the atoms are levitated against gravity to prevent them from falling out of the focus of the vertical imaging system. Relaxation times between 250~ms and 10~s at 250~ms intervals are chosen, with 10 repeats at each time. Each data point comes from a unique experimental run, and relaxation times are randomized to minimize effects from any long-term drifts in experimental conditions. Figure.~\ref{fig:latticeevolution} shows images at several different relaxation times, illustrating the evolution of the system from center-of-mass motion to nucleation of vortices, ordering of vortices into a triangular lattice, and subsequent decay as the angular momentum of the BEC decreases due to friction with the thermal cloud \cite{AboShaeer2002}.


\section{Vortex Detection}

\begin{figure}[b!]
	\centering
		\includegraphics[width=8cm]{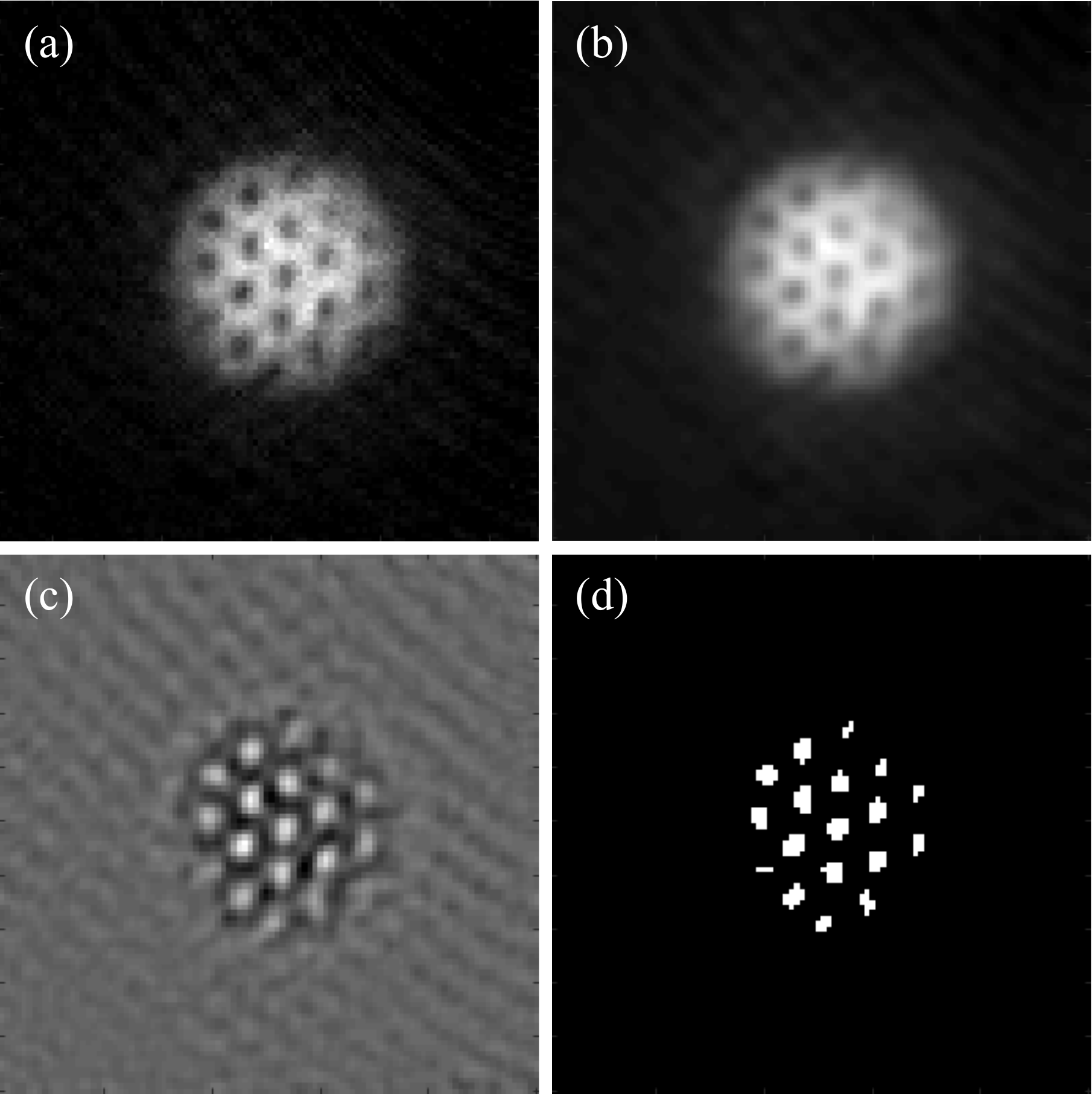}
	\caption{Identifying vortices using a \textit{blob detection} algorithm. The original Fourier filtered image is shown in (a). After applying the convolution with a Gaussian, we get a scale space image (b), where noisy features of the original image are suppressed. (c) The result of applying the Laplacian operator to (b), which emphasizes the vortex cores. (d) The final binary image produced after applying an amplitude $A_{\mathrm{th}}$ threshold to (c). Images shown are 100~$\times$~100 pixels, where each pixel is equal to 5.1~$\times$~5.1~$\mu$m$^2$.}
	\label{fig:blobdetection}
\end{figure}

To extract useful data about vortex number and distribution, we employ a Laplacian-of-Gaussian \textit{blob detection} algorithm \cite{Lindeberg1993}, a technique commonly used in computer vision applications, to automatically locate vortices in each image. We first Fourier filter a given absorption image to remove interference fringes from the probe light. The filtered image is shown in Fig.~\ref{fig:blobdetection}(a). Then by convolving the image $f(x,y)$ with a Gaussian kernel $g(x,y)$ with a width set by the vortex core size in the expanded cloud, we obtain a \textit{scale space} representation \cite{Koenderink1984} [Fig.~\ref{fig:blobdetection}(b)] of our absorption image:
\begin{equation}
L(x,y) = g(x,y) \ast f(x,y).
\label{eq:convolution}
\end{equation} 
This suppresses features smaller than our chosen scale. $g(x,y)$ is simply
\begin{equation}
g(x,y) = \frac{1}{2{\pi}s^2}e^{-\frac{x^2 + y^2}{2s^2}},
\label{eq:gaussian}
\end{equation} 
where $s$ is the width or \textit{scale} of the objects we would like to detect. In our case, these are vortex cores with a Gaussian width of approximately 6~$\mu$m. A Laplacian of this convolution produces positive features for intensity minima, which correspond to the vortex cores [Fig.~\ref{fig:blobdetection}(c)]. Using an amplitude threshold $A_{\mathrm{th}}$, the resulting image is converted to a binary image [Fig.~\ref{fig:blobdetection}(d)], and a size threshold of $p_{\mathrm{th}}$ contiguous pixels is then applied to discard small features that originate from, e.g., spurious dark pixels resulting from imaging artifacts. The vortex positions are finally extracted by determining the ``center of mass'' of each collection of contiguous pixels.

We set $s$ to 1.2 pixels, which is approximately the Gaussian width of a vortex, and set $p_{\mathrm{th}}$ to 3 pixels. $A_{\mathrm{th}}$ is chosen based on the peak values in Fig.~\ref{fig:blobdetection}(c). Varying $s$ and $A_{\mathrm{th}}$ by less than 5$\%$ results in a variation of $\pm$1 detected vortex in a given image. Changing $p_{\mathrm{th}}$ by $\pm$1 pixels results in a variation of $\pm$2 vortices in images with higher numbers of vortices and has less of an effect on images with fewer vortices. False positive detected vortices stem primarily from spurious dark pixels, and they considerably affect subsequent measurements of disorder. We deliberately choose detection parameters that are more likely to undercount vortices than to detect false positives for this reason. The detection algorithm ultimately performs well for images with good contrast, but struggles with detecting vortices in cases where the contrast is reduced, such as near the edge of a cloud or if vortices are not perfectly aligned along the vertical imaging axis. We note that the same detection parameters are applied to the entire data set.


\section{Measuring Disorder}

\begin{figure}
	\centering
		\includegraphics[width=8cm]{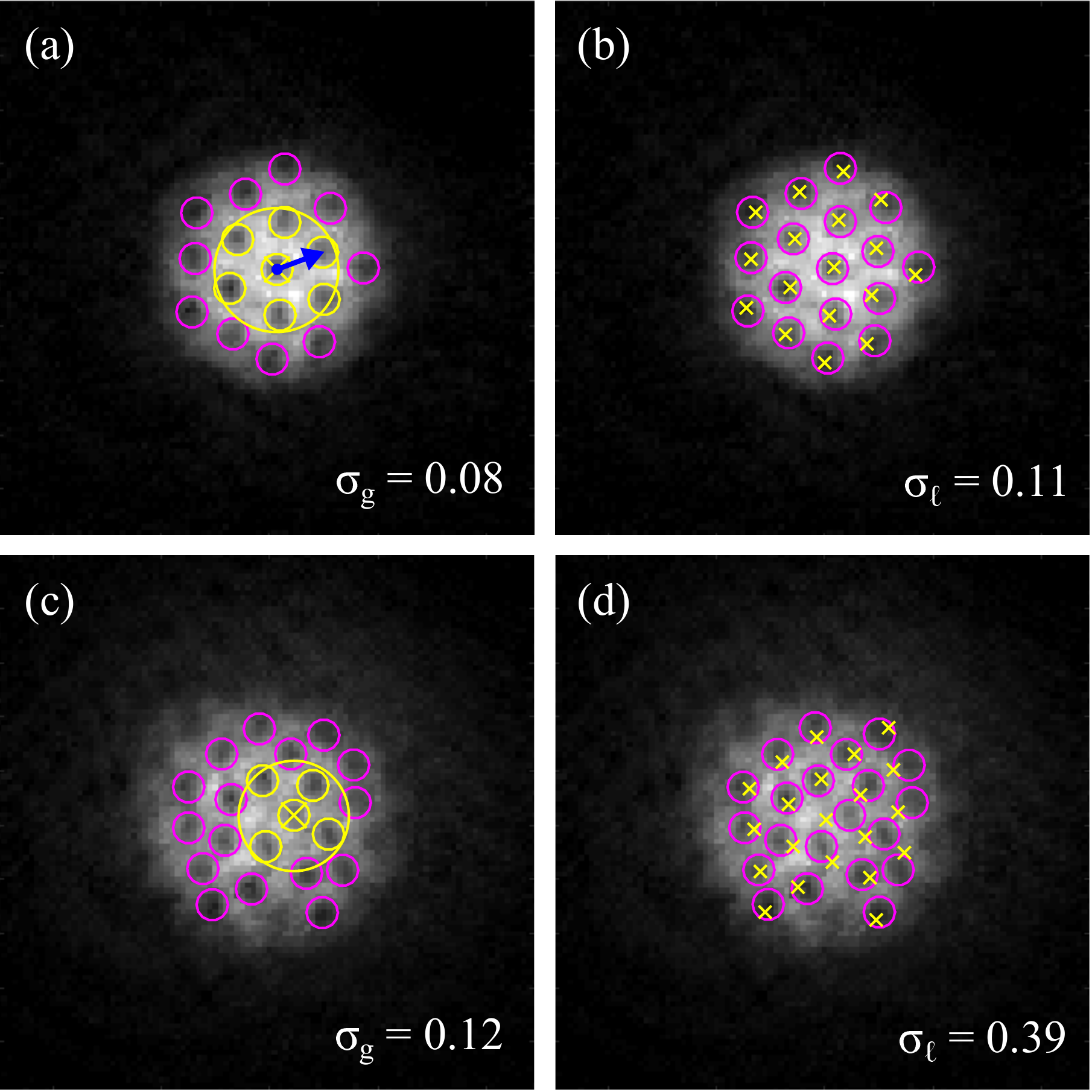}
	\caption{(Color online) Illustration of the two methods used for measuring disorder in an ordered vortex array (top row) and a disordered vortex distribution (bottom row), with algorithmically detected vortices circled in magenta. Each image is 510~$\times$~510~$\mu$m$^2$. After an initial estimate of the average nearest-neighbor spacing $\mu_{\mathrm{ini}}$ based on the vortex distribution near the center of the cloud, the nearest-neighbor search algorithm locates each nearest neighbour for each vortex within a radius specified by $r$, as indicated by the large yellow circles in (a) and (c). In an ordered lattice, such a circle should contain seven evenly spaced vortices, which is clearly not the case for the disordered distribution in (c). In the second method, a triangular lattice is fitted to the vortex distribution. The fitted lattices in (b) and (d) are indicated by the yellow crosses.}
	\label{fig:measuringdisorder}
\end{figure}

We take two different approaches to quantify disorder that are minimally sensitive to imperfections in vortex detection. The first approach makes use of the fact that in a triangular lattice, the nearest neighbors of a given lattice site are equidistant from it. We define the geometric disorder as $\sigma_g = \sigma_{\mathrm{nn}}/\mu_{\mathrm{nn}}$ \cite{Varon2013}, where $\sigma_{\mathrm{nn}}$ is the standard deviation of nearest neighbour distances and $\mu_{\mathrm{nn}}$ is the mean nearest-neighbor distance. In a perfectly ordered triangular lattice, $\sigma_g = 0$. Using a nearest-neighbors search algorithm,\footnote{We used functions available in the \textsc{Matlab} Statistics and Machine Learning Toolbox.} we first analyze the distribution of up to seven vortices nearest to the center of the cloud and use this to estimate the average nearest neighbor-distance $\mu_{\mathrm{ini}}$. This is then used to define a search radius $r = 1.5\mu_{\mathrm{ini}} \sin{(\pi/3)}$ for the full nearest-neighbor search, where $\mu_{\mathrm{ini}}\sin{(\pi/3)}$ is the height of an equilateral triangle with sides of length $\mu_{\mathrm{ini}}$, such that the radius $r$ falls midway between the nearest and next-nearest neighbor distance for a perfect triangular lattice. For each vortex in a cloud, the distances between it and all other vortices within $r$ are found, allowing us to determine $\sigma_{\mathrm{nn}}$ and $\mu_{\mathrm{nn}}$. Each pair of vortices is only counted once. In Figs.~\ref{fig:measuringdisorder}(a) and (c), the nearest neighbors of a vortex marked with an $\times$ within radius $r$ (large yellow circle) are highlighted in yellow.

\begin{figure*}
	\centering
		\includegraphics[width=13cm]{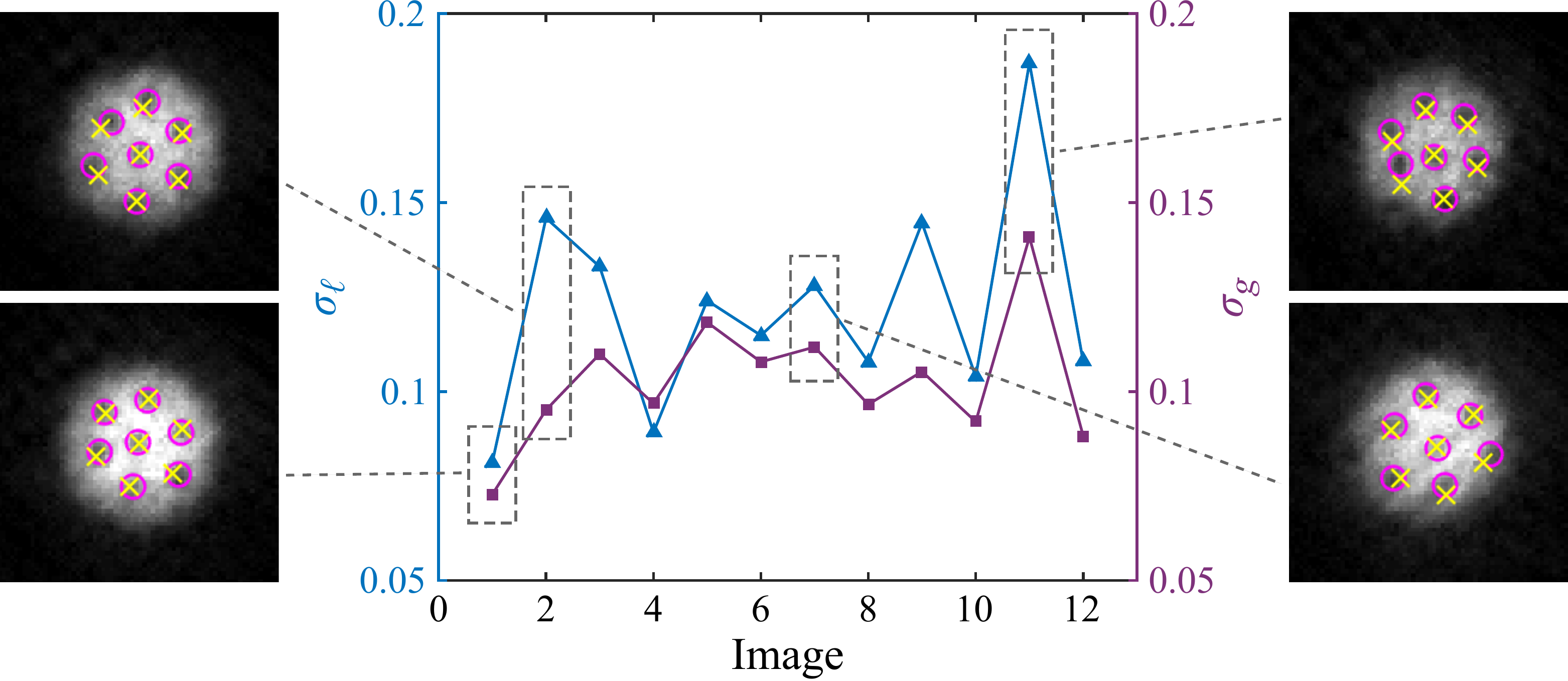}
	\caption{(Color online) Comparing measured disorder of twelve images that each have seven vortices arranged in a hexagon. The left $y$-axis (blue triangles) depicts the lattice disorder $\sigma_{\ell}$. The right $y$-axis (purple squares) depicts the geometric disorder $\sigma_g$. The two methods qualitatively agree with one another. Sample images for four of the points are shown, providing visual corroboration for the measured values of disorder.}
	\label{fig:unithexdisorderexpt}
\end{figure*}

The second approach to determining how much disorder is present in a vortex distribution is to fit a triangular lattice to the distribution. The fitting parameters are the lattice spacing $a$, orientation $\theta$, and the vertical and horizontal position offsets from the cloud center. We make initial guesses of $\theta$ and the position offsets by analyzing the vortex distribution near the center of the cloud. The initial guess for $a$ is set to $\mu_\mathrm{nn}$, since this provides a better initial guess than using only the vortices near the cloud center. This can be seen from Fig.~\ref{fig:measuringdisorder}(c), where it is obvious that the inter-vortex separation near the cloud center is larger than that of the surrounding vortices. We fit a lattice with the same number of lattice sites as vortices using a least-squares fit. Examples of fitted lattices can be seen in Figs.~\ref{fig:measuringdisorder}(b) and \ref{fig:measuringdisorder}(d). Using this approach, the disorder $\sigma_\ell$ is defined in terms of the position deviation of vortices from their corresponding fitted lattice sites:
\begin{equation}
\sigma_\ell = \frac{1}{a}\sqrt{\frac{1}{N}\sum\limits_{i = 1}^N (X_i - x_i)^2 + (Y_i - y_i)^2},
\label{eq:latticedisorder}
\end{equation} 
where $N$ is the number of vortices, $X_i$ and $Y_i$ are coordinates of the fitted lattice sites and $x_i$ and $y_i$ are the coordinates of the vortices. The equation amounts to the root-mean-square average of the position deviation normalized by the lattice spacing and is effectively the cost function of the fit. Similarly to the geometric disorder, a perfectly ordered triangular lattice would have $\sigma_\ell = 0$.

\begin{figure}[!hp]
	\centering
		\includegraphics[width=8.5cm]{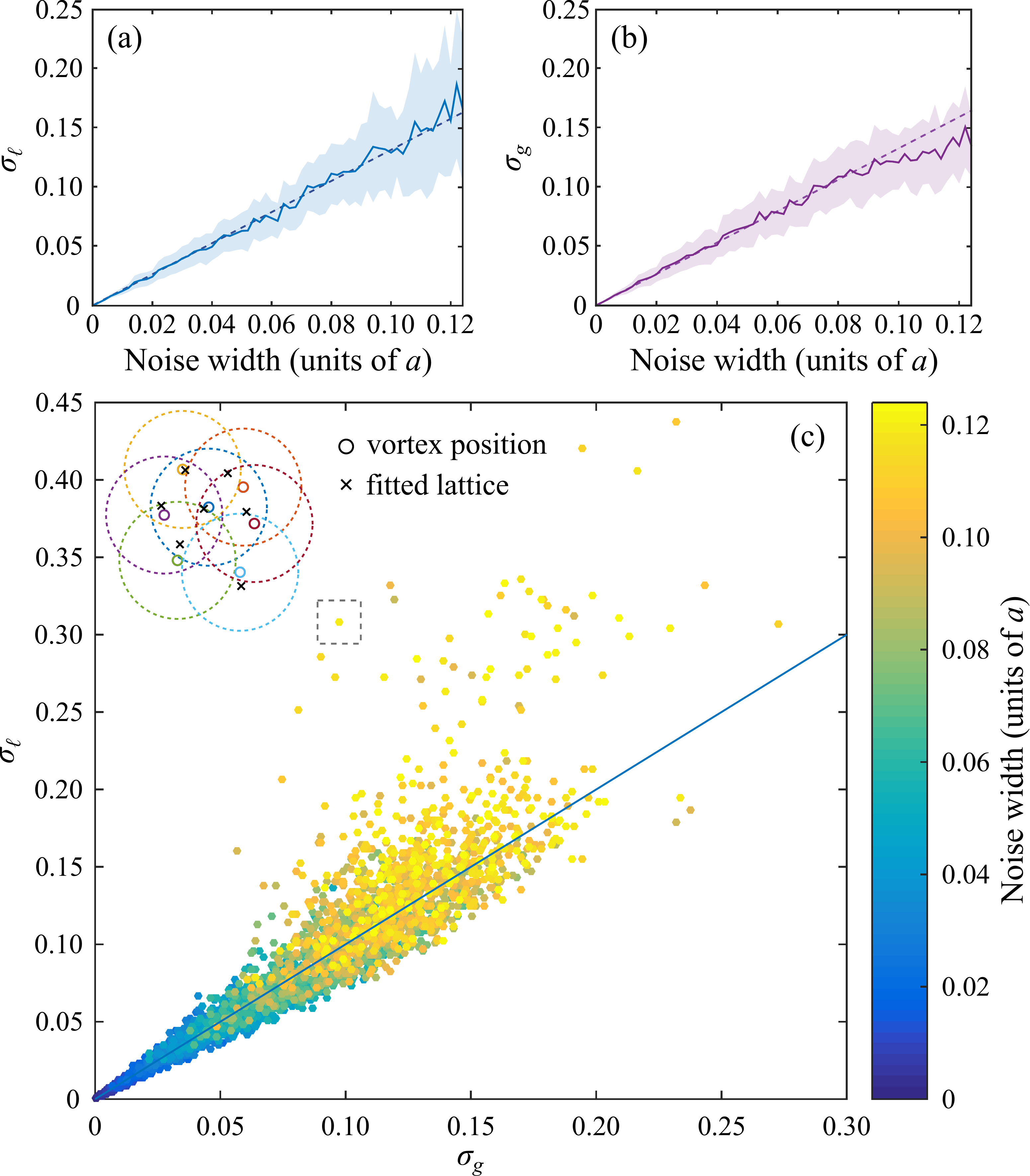}
	\caption{(Color online) Measured disorder of generated vortex distributions of seven vortices with added Gaussian position noise. (a) and (b) show the values of $\sigma_{\ell}$ and $\sigma_g$, respectively, as a function of the width of the Gaussian noise distribution. The solid lines are mean values of 50 theoretical vortex distributions for each value of noise amplitude and the shaded regions in each plot represent the standard deviation. The dashed lines are guides to the eye to illustrate the linear relationship between measured disorder and noise amplitude. $\sigma_{\ell}$ is plotted against $\sigma_g$ in (c). The solid line indicates a slope of unity. As disorder increases, the correlation between $\sigma_{\ell}$ and $\sigma_g$ decreases. An example of a vortex distribution corresponding to the point enclosed by the square in (c), in which $\sigma_{\ell}$ and $\sigma_g$ do not agree, is shown.}
	\label{fig:unithexdisordertheory}
\end{figure}

As a further demonstration of the techniques, we compare all images with seven vortices arranged roughly in a hexagon. As shown in Fig.~\ref{fig:unithexdisorderexpt}, both methods qualitatively produce similar values of disorder and are able to distinguish between a well-ordered vortex lattice and one with defects. To put the disorder values into context, we calculate $\sigma_g$ and $\sigma_\ell$ for theoretical vortex distributions of seven vortices making up a unit hexagon with added Gaussian position noise. As can be seen in Figs.~\ref{fig:unithexdisordertheory}(a) and (b), the disorder grows linearly with noise amplitude in the low noise case, but aspects of each method cause deviations from this trend. For example, when calculating geometric disorder, if a given vortex is sufficiently far away from its nearest neighbors, it will no longer fall within $r$, lowering the value of $\sigma_g$. This can be observed in Fig.~\ref{fig:unithexdisordertheory}(b), where $\sigma_g$ deviates from the linear relationship with the width of the Gaussian noise distribution. Due to the relatively small number of vortices present, such an out of place vortex has a large influence on $\sigma_g$. $\sigma_\ell$ can be sensitive to the initial guess parameters for the least squares fitting routine, which can also converge to a local minimum instead of the global minimum, producing higher values of $\sigma_\ell$, particularly in the case of more disordered vortex distributions.

To determine how $\sigma_\ell$ and $\sigma_g$ are correlated, we plot one as a function of the other in Fig.~\ref{fig:unithexdisordertheory}(c). The correlation between the two measures is good for the bulk of the points, but significant deviations occur for larger noise widths. An example of a vortex distribution that leads to such a disparity is shown in the figure, where some vortices are ``missed'' because they are not enclosed by $r$, giving a lower than expected $\sigma_g$. This is not expected to be as much of a problem in systems where many more vortices are present.

Despite the flaws of each individual method, the two complement one another. For instance, the geometric disorder value is not very sensitive to lattice dislocations, whereas a dislocation leads to high values of $\sigma_\ell$. In Fig.~\ref{fig:measuringdisorder}(d), one can see that the vortices in the upper left of the cloud are ordered while the rest are not, leading to a moderate $\sigma_g$, but a high $\sigma_\ell$. Selecting images with very different values of disorder for reanalysis could be used to automatically detect features such as grain boundaries in systems with many vortices.

\begin{figure}
	\centering
		\includegraphics[width=8.5cm]{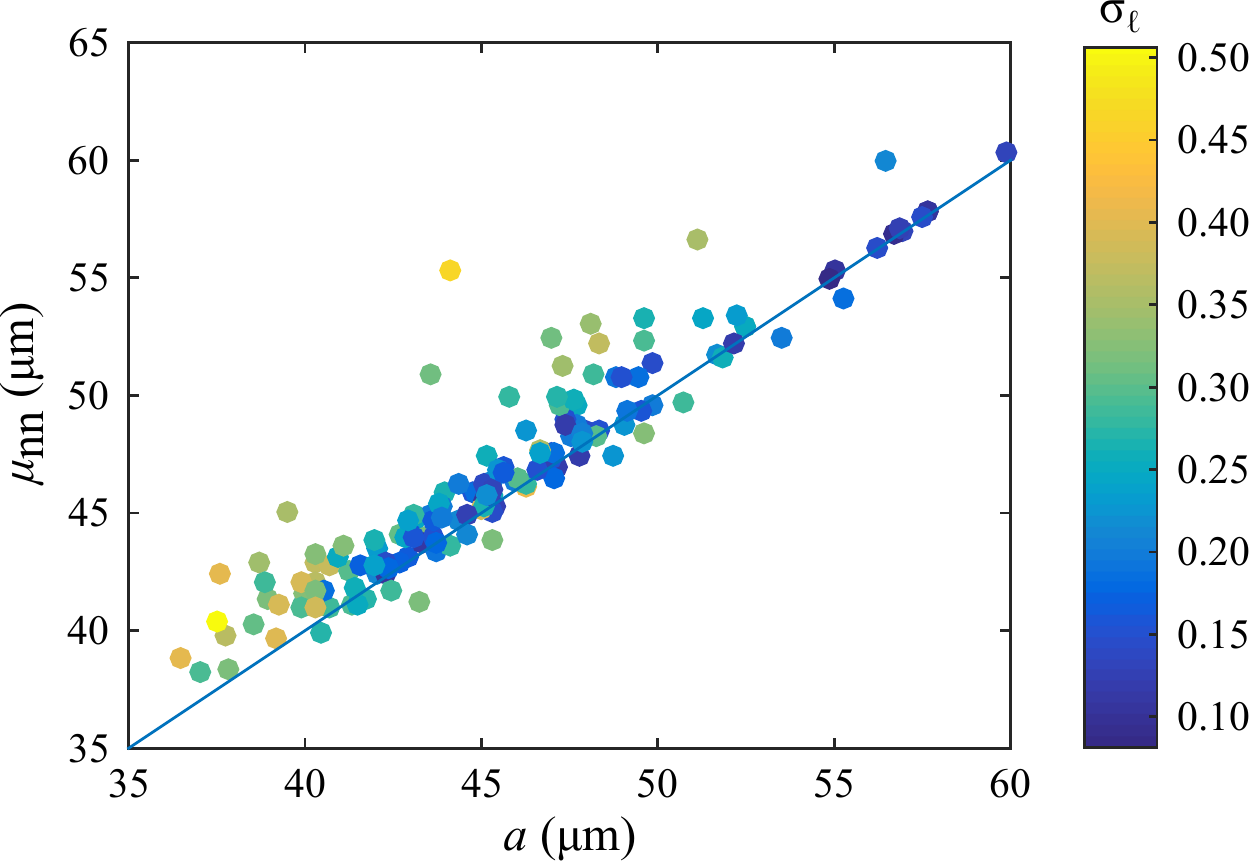}
	\caption{(Color online) Mean nearest-neighbor distance, $\mu_{nn}$, plotted against the fitted lattice spacing, $a$, for all images with seven or more vortices. Ideally, one would expect a slope of unity (solid line). Deviations from the ideal slope increase as a function of $\sigma_\ell$, indicating a poor lattice fit.}
	\label{fig:munnvsa}
\end{figure}

In Fig.~\ref{fig:munnvsa}, we plot $\mu_{\mathrm{nn}}$ against $a$ for all images with seven or more vortices. Images with fewer vortices are not included because fewer than seven vortices are not expected to form an equilateral triangular lattice. In the initial time after stirring, many vortices are present in the system and are consequently relatively densely packed as well as being disordered. This is evident in the short length scale end of the plot. When the vortices are ordered in a perfect lattice, $\mu_{\mathrm{nn}}$ and $a$ should be equal. Indeed, this is the case; as disorder increases, the lattice fit becomes less representative of the actual vortex distribution, and $a$ deviates from $\mu_{\mathrm{nn}}$. 

\section{Vortex Evolution}

\begin{figure}
	\centering
		\includegraphics[width=8.5cm]{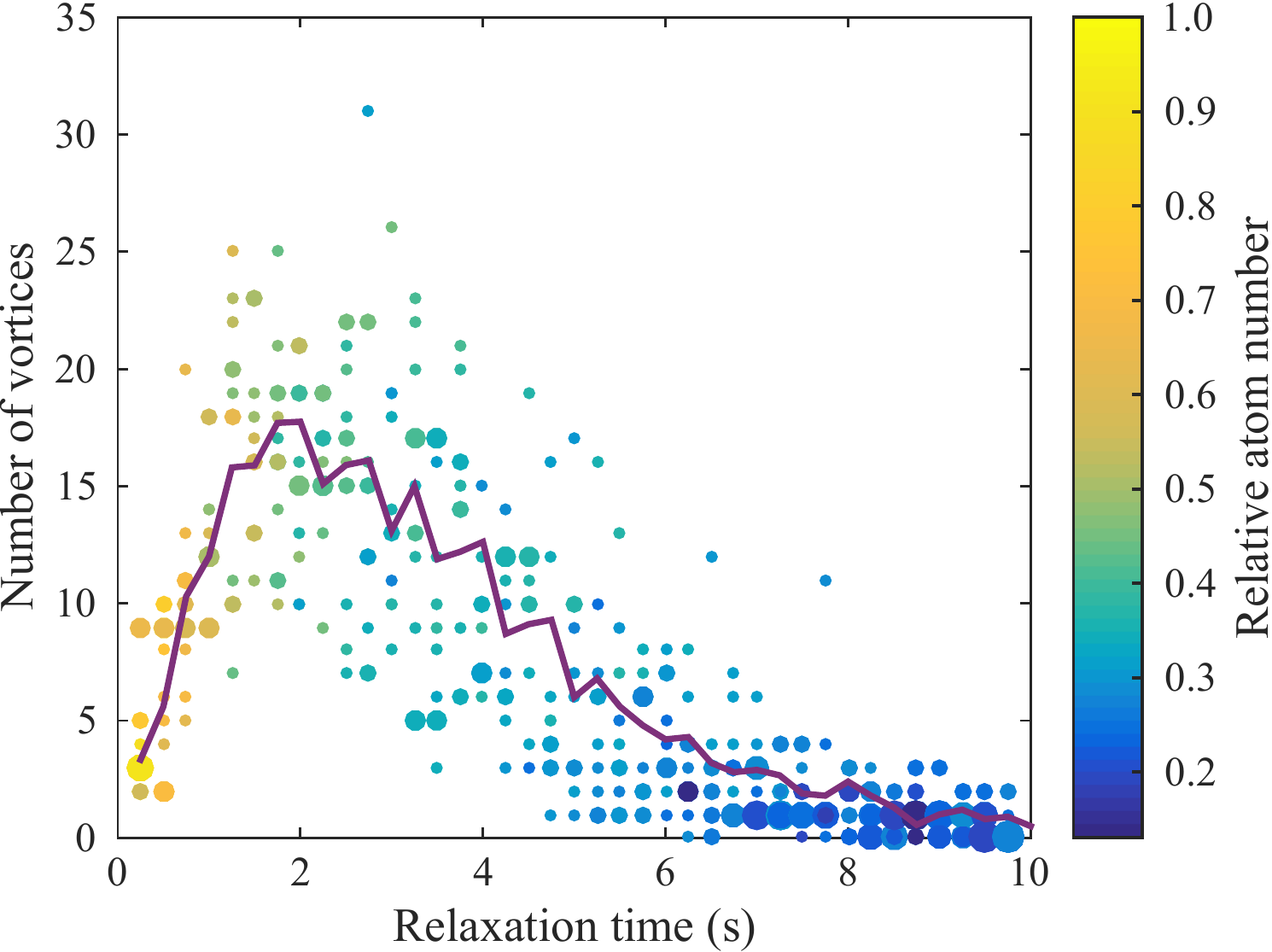}
	\caption{(Color online) Number of vortices as a function of relaxation time after imparting rotation. The color scale indicates the relative number of atoms, scaled to an initial number of 8.3 $\times~10^5$ prior to rotation, which predictably decreases over time. The size of each point indicates frequency of occurrence, with larger points corresponding to more experimental runs with the same number of vortices. The solid line is the mean number of vortices at each time.}
	\label{fig:vorticesvstime}
\end{figure}

\begin{figure}[b]
	\centering
		\includegraphics[width=8.5cm]{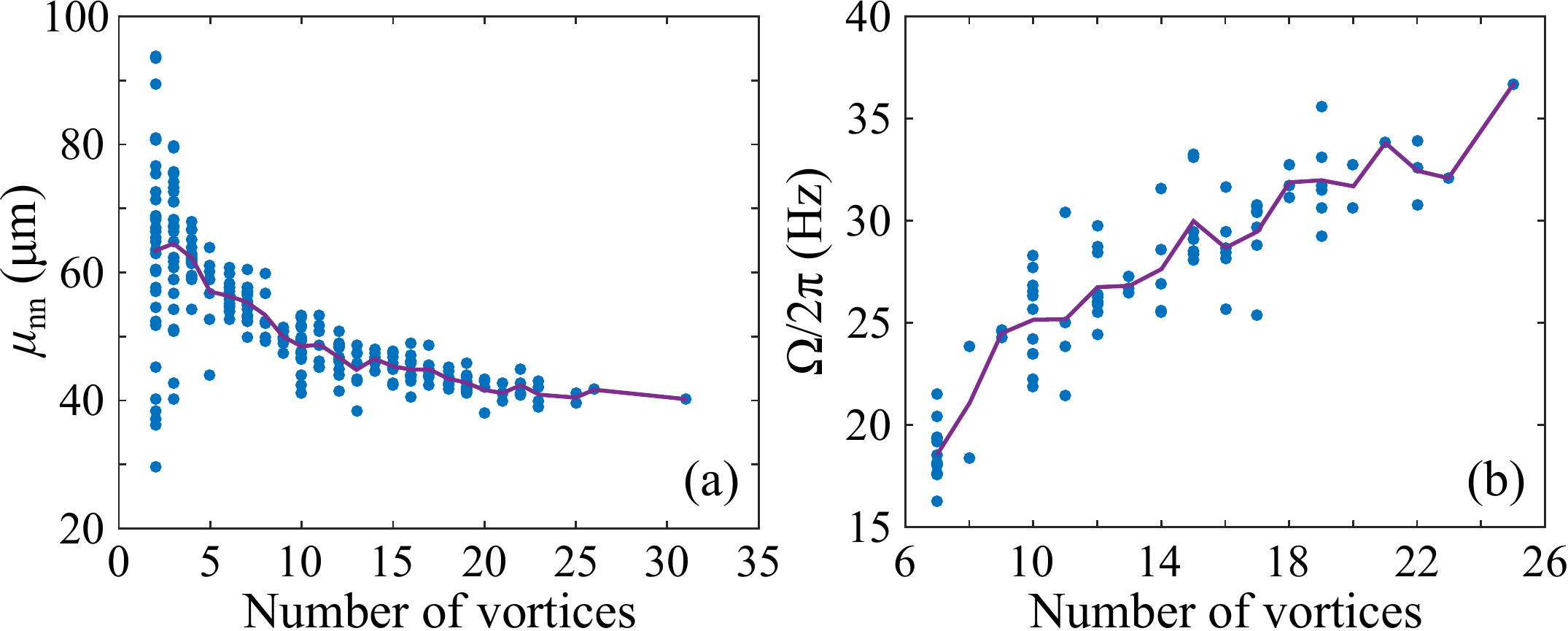}
	\caption{(Color online) (a) $\mu_{nn}$ as a function of number of vortices. As expected, the vortex spacing is inversely proportional to the number of vortices present. (b) Angular speed $\Omega$ as a function of number of vortices for images where an ordered vortex lattice is present. Note that we typically underestimate vortex number, which is a significant source of variance in both plots. The solid lines are the mean values.}
	\label{fig:daveplots}
\end{figure}

We now apply the vortex detection and ordering analysis to examine the evolution of the vortex lattices in the experiment. Vortex number data as a function of relaxation time for the entire data set is plotted in Fig.~\ref{fig:vorticesvstime}, along with relative atom number on the color scale. A small number of images were not included in the analysis due to ambiguous vortex detection as a result of, e.g., vortices not aligned with the imaging axis. It is clear that there is a large variation of vortex number at a given relaxation time. The number of vortices is only weakly correlated with atom number and is not correlated to condensate fraction, which rules out the effects of shot-to-shot repeatability of atom number and temperature in the experiment. Instead, we attribute the variation to a combination of two potential causes. The first is a long term instability in the ambient magnetic field, which can cause variations in both the quadrupole field zero position and the trapping frequency. This is suggested by the irreproducibility of the center-of-mass position of the condensate from run to run. The second potential cause is residual center-of-mass motion of the condensate coupling angular momentum back into the cloud, effectively causing a ``revival'' of the vortex lattice in some cases, such as the point with 12 vortices at 7.75~s. Theoretical modelling is required for further investigation.

\begin{figure}
	\centering
		\includegraphics[width=8.5cm]{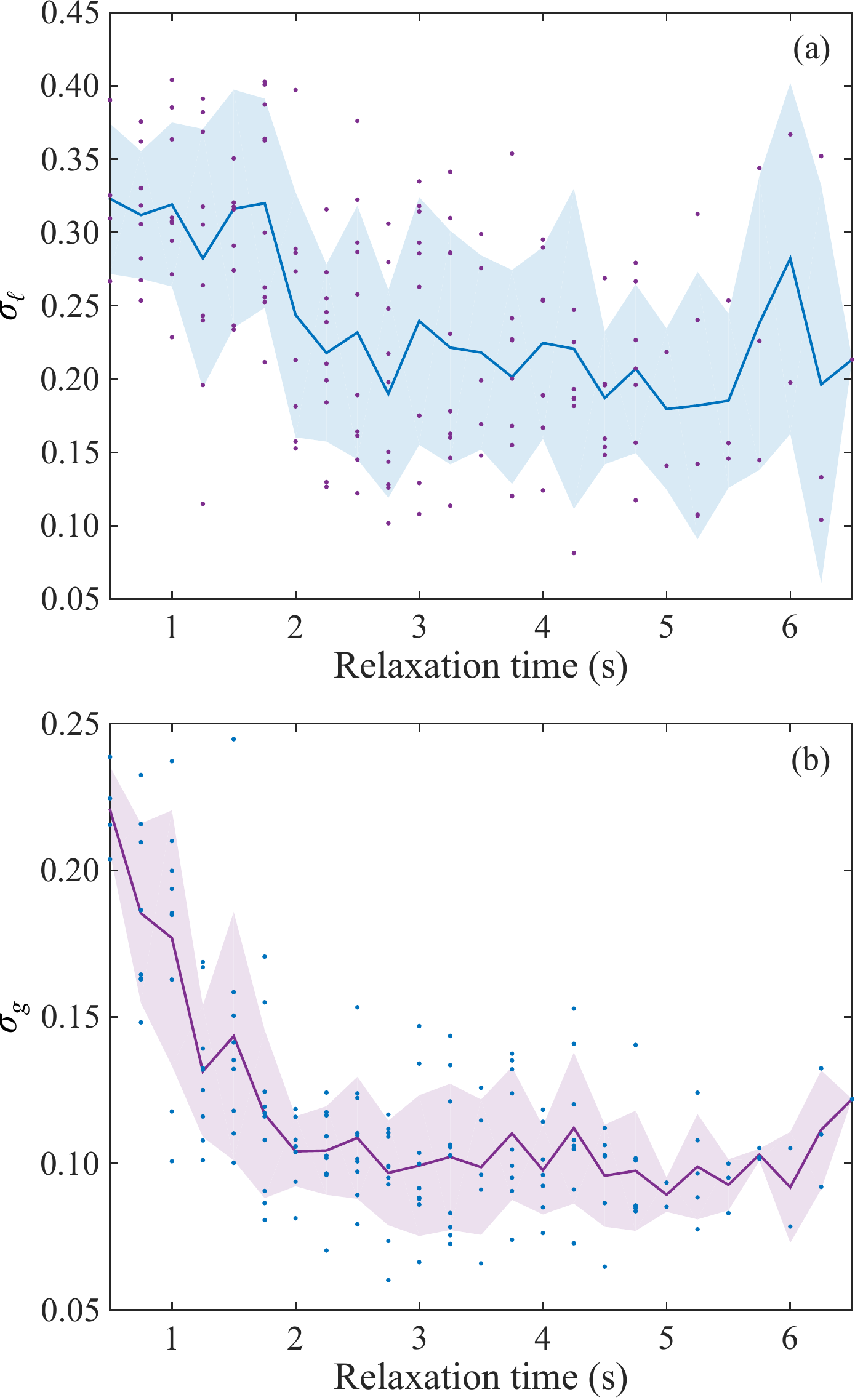}
	\caption{(Color online) (a) Lattice disorder and (b) geometric disorder as a function of relaxation time for images with at least seven vortices, showing that the lattice becomes ordered after 2~s.}
	\label{fig:disordervstime}
\end{figure}

With higher rotational velocity of the condensate, the vortex density increases \cite{Sheehy2004, Coddington2004}. We can see this in Fig.~\ref{fig:daveplots}(a), where $\mu_{\mathrm{nn}}$ is plotted against the number of vortices for all images with at least two vortices and at relaxation times longer than 1~s, when the substantial center-of-mass motion of the cloud has died down. In the case of fewer than seven vortices being present, the equilibrium configuration is not a triangular lattice. If a uniform ordered lattice is present, we can express the vortex density $n_v$ as a function of $a$ \cite{Coddington2004}:
\begin{equation}
n_v = \frac{2}{\sqrt{3}a^2}.
\label{eq:vortexdensity}
\end{equation} 
If we assume only rigid body rotation, that is $n_v = {\Omega}m/{\pi}{\hbar}$, where $m$ is the atomic mass and $\Omega$ is the angular velocity, we can calculate $\Omega$ as a function of $a$ independently of the number of vortices. In Fig.~\ref{fig:daveplots}(b), we plot $\Omega$ as a function of the number of vortices for images with seven or more vortices at relaxation times greater than 1~s and $\sigma_\ell < 0.25$, where an ordered is lattice present. We find that for the highest number of vortices in this category, $\Omega$ is similar to the driving frequency of 37~Hz.

To study how the order of the vortex lattice evolves over time quantitatively, we calculate $\sigma_g$ and $\sigma_\ell$ for the entire data set. Figure~\ref{fig:disordervstime} shows the evolution of order using each metric. Using both measures, a clear change in the mean values can be seen at 2~s. This agrees with the visual observation that the lattice crystallizes at around this time. We note that this coincides with the time when the maximum number of vortices is detected in the cloud. The stirring process leaves the condensate in a highly non-equilibrium state with numerous vortices entering the system at the edge of the condensate. Not all of these vortices are detected due to the low density in this region. However, after approximately 2~s, the dynamics between the condensate and thermal cloud have largely settled; new vortices are no longer being added to the system and an ordered lattice can therefore form.

While $\sigma_g$ steadily increases with decreasing relaxation time, $\sigma_\ell$ appears to level off at just above a value of 0.3. This limit can be understood from a simple geometric argument. Suppose that our vortices are arranged in a perfect triangular lattice with lattice spacing $a$. If one vortex is displaced such that it is maximally distant from any surrounding lattice sites, i.e., it is at the center of an equilateral triangle, its distance from the nearest lattice site is $d_{\mathrm{max}} = \sqrt{7}a/4 \sim0.66a$. We can suppose that a vortex lattice is maximally disordered when half of the vortices are displaced by $d_{\mathrm{max}}$ from the ideal lattice sites. A fit to such a vortex distribution would result in each vortex being $d_{\mathrm{max}}/2$ away from a given fitted lattice site. From Eq.~(\ref{eq:latticedisorder}), this results in $\sigma_\ell^{\mathrm{max}} = d_{\mathrm{max}}/2a \sim0.33$.

The effective upper limit on $\sigma_\ell$ makes it a poor metric for distinguishing how disordered a vortex lattice is when the disorder is high. Nevertheless, the transition between an ordered and disordered lattice is clear. On the other hand, $\sigma_g$ can identify the continuous increase in order throughout the crystallization process.


\section{Discussion and Outlook}

We have reported the observation of vortex dynamics in a BEC confined in a hybrid optical-magnetic trap. Vortices were induced by rotating the anharmonic magnetic potential that provides confinement in the horizontal plane, and evolution of a vortex lattice was studied by analyzing vortex distributions at a range of relaxation times. We have described a method to automatically detect vortices using a blob detection algorithm. We then applied and evaluated two measures of disorder in the vortex lattice, one based on the spread of nearest-neighbor distances and the other derived from fitting a triangular lattice to the vortex distribution. Using these methods, we were able to straightforwardly extract information about the vortex distribution and use it to estimate the rotational velocity of the condensate. We have shown that both $\sigma_g$ and $\sigma_\ell$ are able to distinguish between ordered and disordered vortex distributions, and that $\sigma_g$ can be used to track the crystallization of a triangular vortex lattice.

By using a method that relies purely on image analysis, we can provide a heuristic approach to measure the lattice energy via the calculated disorder. Future theoretical work will investigate the relationship between disorder and the vortex lattice energy in our system. Experimentally, we will employ the techniques developed here to study how the lattice crystallization time varies with experimental parameters, such as temperature and trap frequencies. An additional advantage of image analysis-based methods is that they can be applied to systems that exhibit ordering beyond only cold atoms.

To study lattice evolution more carefully, we must overcome the shot-to-shot irreproducibility of vortex number at a given relaxation time that arises from residual motion of the condensate and/or slow variations in trap parameters. Better magnetic field stability would address the latter problem, but the nature of the stirring mechanism limits our ability to control residual center-of-mass motion of the BEC. To fully understand lattice formation and decay dynamics, it will ultimately be necessary to implement a minimally-destructive \textit{in situ} imaging technique, such as Faraday imaging \cite{Gajdacz2013} or partial transfer absorption imaging \cite{Ramanathan2012}, to complement images of the expanded cloud. Future experiments in different trap geometries, such as quasi-2D and more anisotropic 3D geometries, combined with 3D Gross-Pitaevskii equation modeling may shed light on energy dissipation mechanisms that allow lattice crystallization.

The data presented in this paper are available for download \cite{SupportingData}.

\section{Acknowledgments}

We acknowledge the UK Engineering and Physical Sciences Research Council (Grants No.~EP/L010844/1 and No.~EP/K030558/1) for funding. T.P.B.~acknowledges financial support from the John Templeton Foundation via the Durham Emergence Project (\url{http://www.dur.ac.uk/emergence}).


%

\end{document}